\documentclass[onecolumn,final]{elsart3p}

\bibstyle{elsart-num}

\usepackage{amssymb}

\journal{Physica A}

\begin{document}
\begin{frontmatter}

\title{Molecular random walks in a fluid and an invariance group \\ of
the Bogolyubov generating functional equation}

\author{Yuriy E. Kuzovlev}
\ead{kuzovlev@kinetic.ac.donetsk.ua}
\address{Donetsk A.\,A.\,Galkin Institute for Physics and Technology
of NASU, 83114 Donetsk, Ukraine}


\begin{abstract}
The problem of statistics of molecular random walks in a classical
fluid is analyzed by means of the BBGKY hierarchy of equations
reformulated in terms of the Bogolyubov evolution equation for
generating functional of many-particle distribution functions. A
proper equivalent set of correlation functions is introduced so that
all they are integrable, vanish in statistical equilibrium, otherwise
accumulate statistical information about history of collisions of a
``molecular Brownian particle'' (test molecule) with other molecules
of the fluid. An exact evolution equation for generating functional
of such correlation functions is derived. Then it is shown that
time-dependent solution to this equation, as well as a properly
defined generating functional of static thermodynamically equilibrium
correlations, possesses invariance with respect to a definite group
of transformations of independent variables of the functional, if
density of the fluid (number of molecules per unit volume) is treated
as one of the independent variables. Such invariance results in
infinitely many exact relations between the correlation functions and
probability distribution of path of the molecular Brownian particle.
Even simplest of these relations suggest significant restrictions on
a profile of the path probability distribution, even without literal
solving the BBGKY hierarchy.
\end{abstract}


\end{frontmatter}

\section{Introduction}

Everybody knows that all thermodynamic and kinetic properties of a
fluid are determined by thermal motion of its atoms or molecules.
This motion by its nature is unbound and represents a ``random walk''
or, in other words, ``Brownian motion''. What kind of statistics
characterizes it?

Strangely enough, this principally interesting question never was a
special issue of those fundamental approach to statistical mechanics
of fluids what originated from the Bogolyubov's book \cite{bog} and
resulted in the Bogolyubov-Born-Green-Kirkwood-Yvon (BBGKY) infinite
hierarchy of equations \cite{bog,uf,re,bal,dor}. Unfortunately, next
work of Bogolyubov and other theorists was concentrated at not
conscientious solving these equations but forcible reducing them to a
single ``kinetic equation'' for one-particle distribution function,
at that being guided by the old Boltzmann's ``molecular chaos''
hypothesis. The latter assumes that two molecules entering into
collision are statistically independent one on another, at least in
case of dilute gas.

Such hypotheses rise up from naive identifying dynamical independence
of particles and statistical one, in spite of that the first concerns
individual phase trajectory of a dynamical system while the second
statistical ensemble of trajectories corresponding to some ensemble
of their initial conditions. But Krylov in his remarkable book
\cite{kr} (first published in 1950) finally proved that dynamical
independence gives no grounds for statistical one. Although, to tell
the truth, this is clear as it is. Analogously, in a statistical
ensemble of the mankind histories any two contemporaries would have
somehow statistically dependent destinies even if they reside in
different continents and do not influence one another. All the more
if they do. Saying about gases, for example, in \cite{i1} (see also
\cite{i2}) it was shown that in spatially non-uniform statistical
ensemble the pair distribution function for atoms in pre-collision
states does not reduce to one-particle distribution function, even
under the limit of infinitely dilute gas (or so-called Boltzmann-Grad
limit).

In fact, fortunately, the BBGKY hierarchy is self-consistent and has
no need in extraneous hypotheses. It is quite another matter if a
technique of its correct solving is in embryo. In respect to the
above formulated question, an approximate approach was suggested in
\cite{i1} (or see \cite{i2}). Its results were principally different
from predictions of conventional theory operating with the Boltzmann
and Boltzmann-Lorentz or similar equations \cite{uf,re,bal,dor}
implied by the ``molecular chaos'' hypothesis. But since then no
response to \cite{i1} was noticed. Hence, importance of the question
still is not realized. This circumstance stimulated work \cite{p1}
where due to simplification and improvement of method of \cite{i1} a
whole probability distribution of Brownian path of a test atom was
approximately found (instead of its four lowest statistical moments
as in \cite{i1}). Recently in \cite{pro} and then in
\cite{jstat,last} a new approach to the same problem was developed
which exploits the ``generalized fluctuation-dissipation relations''
\cite{j12,p}. A peculiar exact ``virial expansion'' for the
probability distribution of Brownian path was obtained which allows
to describe characteristic spatial profile of this distribution
without any quantitative approximations at all. At that, results of
\cite{p1} were confirmed.

Well, clearly, now considerations of the BBGKY hierarchy are next in
turn to be improved. With this purpose, in the present paper we make
use of the Bogolyubov's formulation of the BBGKY hierarchy in terms
of generating functional of many-particle distribution functions and
evolution equation for this functional \cite{bog}. Notice that
Bogolyubov himself analyzed stationary solution to this equation
only, that is mere thermodynamically equilibrium functional, and in
fact had not investigated its non-equilibrium time-dependent
solutions. Apparently, because of this an interesting property of the
generating functional and its evolution equation remained unnoticed,
and we had to catch it.

The matter concerns\, (i) logarithmic variational derivative of the
equilibrium functional,\, (ii) generating functional of properly
introduced non-equilibrium $\,n$-particle ``correlation functions''
which together represent statistical correlations between a total
previous path of the ``molecular Brownian particle'' (test atom) and
current state of the fluid,\, and (iii) time evolution operator for
this new functional. All these three objects turn out to be invariant
with respect to definite continuous group of transformations of their
independent variables. The latter are (i) arbitrary probe function
defined in the 6D $\,\mu$-space and (ii) density of fluid (number of
particles per unit volume).

It will be shown that this invariance property implies an infinite
set of exact relations connecting the probability distribution of
Brownian path and the correlation functions . A solution to the BBGKY
hierarchy automatically satisfy all these relations. One of them
coincides with the mentioned ``virial expansion''. Therefore they
present essential information about solution to the BBGKY hierarchy
and even can prescribe it accurate to quantitative details. Let us go
to these curious things.

\section{Equations of molecular Brownian motion and Bogolyubov's
generating functional}

Let a macroscopic box with volume $\,\Omega\,$ contains $\,N\gg 1\,$
identical atoms plus one more different but similar particle which
will be under our special attention. The atoms have mass $\,m\,$,
coordinates $\,{\bf r}_j\,$ and momenta $\,{\bf p}_j\,$
($\,j=1,2...\,N\,$) and all interact one with another by means of
spherically symmetric potential $\,\Phi_a({\bf r}_j-{\bf
r}_k)=\phi_a(|{\bf r}_j-{\bf r}_k|/r_a)\,$. The additional particle
has mass $\,M\,$, coordinate $\,{\bf R}\,$, momentum $\,{\bf P}\,$
and interacts with all atoms through spherically symmetric potential
$\,\Phi_b({\bf r}_j-{\bf R})=\phi_b(|{\bf r}_j-{\bf R}|/r_b)\,$.  Due
to the interactions, this particle undergoes, along with atoms, a
chaotic wandering, therefore we can name it ``Brownian particle''
(BP). We want to consider probability distribution of its position,
$\,{\bf R}(t)\,$, assuming that at some initial time moment, e.g.
$\,t=0\,$, it is certainly known:\, $\,{\bf R}(t=0)={\bf R}_0\,$.

On the contrary, personal coordinate of any concrete atom all the
time is unknown and possesses nearly uniform probability distribution
over $\,\Omega\,$. In such situation, normalized particular
distribution functions for the BP together with any $\,n=0,1,2,...\,$
atoms, following Bogolyubov \cite{bog}, can be written as
\[
D_n(t,{\bf R},{\bf r}^{(n)},{\bf P},{\bf
p}^{(n)})\,=\,\Omega^{-\,n}\, F_n(t,{\bf R}, {\bf r}^{(n)},{\bf
P},{\bf p}^{(n)}|{\bf R}_0;\nu_{\,0})\,\,\,,
\]
where $\,{\bf r}^{(n)}=\{{\bf r}_1...\,{\bf r}_n\,\}\,$, $\,{\bf
p}^{(n)}=\{{\bf p}_1...\,{\bf p}_n\,\}\,$, and new distribution
functions $\,F_n\,$ depend on $\,\Omega\,$ through mean density
(number of atoms per unit volume), $\,\nu_0=N/\Omega\,$, only.
Argument $\,{\bf R}_0\,$ is added in order to remind of the mentioned
initial condition for BP. At that
\[
F_n(t,{\bf R}, {\bf r}^{(n)},{\bf P},{\bf p}^{(n)}|{\bf
R}_0;\nu_{\,0})\,=\,\Omega^{n}\int_{n+1} ...\int_N D_N(t,{\bf R},
{\bf r}^{(N)},{\bf P},{\bf p}^{(N)})\,\,
\]
with\, $\,\int_k ...=\int\int ...\,\,d{\bf r}_k\,d{\bf p}_k\,$\,\,
and $\,D_N\,$ being normalized probability distribution function of
the whole system. This distribution function (DF) is solution to the
Liouville equation
\[
\frac {\partial D_N}{\partial t}\,=\,[\,L_N\,+\,L^{(box)}_N\,]\,D_N\,
\]
with initial condition, in the framework of the Gibbs canonical
statistical ensemble,
\begin{equation}
\begin{array}{l}
D_N(t=0,...\,)\,=\,\delta ({\bf R}-{\bf R}_0)\,\,e^{-\,H_N/T}\,\left[
\int d{\bf R}\int d{\bf P}\int_1...\int_N \delta ({\bf R}-{\bf
R}_0)\,e^{-\,H_N/T}\,\right ]^{-\,1}\,\,\label{din}
\end{array}
\end{equation}
Symbol\, $\,H_n\,$ hear means Hamiltonian of system ``$\,n\,$ atoms
plus BP in the box'', $\,L_n+L_n^{(box)}\,$ is the Liouville operator
corresponding to $\,H_n\,$, and $\,L_n\,$ is its autonomous
box-independent part:
\[
\widehat{L}_n\,=\,-\frac {\bf P}{M}\cdot\frac {\partial }{\partial
{\bf R}}-\sum_{j\,=1}^n\frac {{\bf p}_j}{m}\cdot\frac {\partial
}{\partial {\bf
r}_j}\,+\sum_{j\,=1}^n\widehat{L}_{j}^{b}\,+\sum_{1\leq j<k \leq n}
\widehat{L}_{jk}^{a}\,\,\,,
\]
where
\[
\widehat{L}_{j}^{b}\,\equiv\,\nabla \Phi_{b}({\bf R}-{\bf
r}_{j})\cdot \left(\frac {\partial }{\partial {\bf P}}-\frac
{\partial }{\partial {\bf p}_j}\right
)\,\,\,,\,\,\,\,\,\,\,\,\widehat{L}_{jk}^{a}\,\equiv\,\nabla
\Phi_{a}({\bf r}_j-{\bf r}_k)\cdot \left(\frac {\partial }{\partial
{\bf p}_j}-\frac {\partial }{\partial {\bf p}_k}\right )\,\,
\]
and\, $\,\nabla\Phi({\bf r})=\partial \Phi({\bf r})/\partial {\bf
r}\,$.

Choosing initial distribution in the form (\ref{din}), we thus
confine ourselves by investigation of random walk of the BP in a
thermodynamically equilibrium fluid.

Under reasonable interaction potentials $\,\phi_{a,\,b}(x)\,$ (fast
enough tending to infinity at $\,x\rightarrow 0\,$ and to zero at
$\,x\rightarrow\infty \,$) the particular DF $\,F_n\,$ have definite
``thermodynamical limit'' when walls of the box withdraw to
infinity,\, $\,N\rightarrow\infty \,$\, and\,
$\,\Omega\rightarrow\infty\,$\, at\, $\,N/\Omega =\nu_{\,0}=
\,$\,\,const\,.\, In this limit the Liouville equation produces
\cite{bog} infinite hierarchy of the BBGKY equations for
$\,F_n\,$\,\cite{uf,re,bal,dor}. In our case it appears as
\begin{equation}
\frac {\partial F_0}{\partial t}\,=\,-\frac {{\bf P}}{M}\cdot\frac
{\partial F_0}{\partial {\bf R}}\, +\nu_0\int_1
\widehat{L}_{1}^{b}\,F_1\,\,\,,\label{f0}
\end{equation}
\begin{equation}
\frac {\partial F_1}{\partial
t}\,=\,\widehat{L}_{1}\,F_1\,+\,\nu_0\int_2
\widehat{L}_2^{b}\,F_2\,+\,\nu_0\int_2\widehat{L}_{12}^{a}\,F_2
\,\,\,,\label{f1}
\end{equation}
\begin{equation}
\frac {\partial F_n}{\partial
t}\,=\,\widehat{L}_{n}\,F_n\,+\,\nu_0\int_{n+1}
\widehat{L}_{n+1}^{b}\,F_{n+1}\,+\,\nu_0
\sum_{j\,=1}^n\,\,\int_{n+1}\widehat{L}_{j\,n+1}^{a}
\,F_{n+1}\,\,\label{fn}
\end{equation}
The only difference of these equations from routine ones
\cite{bog,uf,re,bal,dor} is presence of a distinct particle, namely,
the BP. Integration over its variables eliminates equation (\ref{f0})
and turns (\ref{f1})-(\ref{fn}) to simplest BBGKY hierarchy for a
system of equivalent particles. According to (\ref{din}), initial
conditions for our hierarchy are
\begin{equation}
\begin{array}{l}
F_0(0,{\bf R},{\bf P}|{\bf R}_0;\nu_0)\,=\,\delta({\bf R}-{\bf
R}_0)\,G_M({\bf P}) \,\,\,,\,\label{ic}\\
F_n(0,{\bf R},{\bf r}^{(n)},{\bf P},{\bf p}^{(n)}|{\bf
R}_0;\nu_0)\,=\, \delta({\bf R}-{\bf R}_0)\,F_n^{(eq)}({\bf
r}_1...\,{\bf r}_n|{\bf R};\nu_0)\,G_M({\bf P})\prod_{j\,=1}^n
G_m({\bf p}_j)\,\,\,,
\end{array}
\end{equation}
where
\[
\begin{array}{l}
G_m({\bf p})\,=\,(2\pi Tm)^{-\,3/2}\exp{(-{\bf p}^2/2Tm)}\,\,
\end{array}
\]
is Maxwell momentum distribution of a particle with mass $\,m\,$,\,
and $\,F_n^{(eq)}({\bf r}_1...\,{\bf r}_n|{\bf R};\nu_0)\,$ is usual
thermodynamically equilibrium joint DF of coordinates of $\,n\,$
atoms and BP corresponding to the canonical ensemble. Of course,
eventually we are most interested in evolution of $\,F_0(t,{\bf
R},{\bf P}|{\bf R}_0;\nu_{\,0})\,$ and especially in probability
distribution of BP'path,
\[
V_0(t,\Delta{\bf R};\nu_{\,0})\,=\,\int F_0(t,{\bf R},{\bf P}|{\bf
R}_0;\nu_{\,0})\,d{\bf P}\,\,\,\,\,\,\,\,\,(\,\Delta{\bf
R}\,\equiv\,{\bf R}-{\bf R}_0\,)\,
\]

It is important that $\,F_n\,$ by their definition naturally satisfy
weakening of correlations between distant particles. Under initial
conditions (\ref{din}) and (\ref{ic}) this property concretizes to
\begin{equation}
\begin{array}{l}
F_1(t,{\bf R},{\bf r}_1,{\bf P},{\bf p}_1 |{\bf
R}_0;\nu_0)\,\rightarrow\, F_0(t,{\bf R},{\bf P}|{\bf
R}_0;\nu_0)\,G_m({\bf p}_1)\,\,\,,\label{unc}\\
F_1^{(eq)}({\bf r}_1|{\bf R}_0;\nu_0)\,\rightarrow\,1\,\,\,,\\
F_n(...\,{\bf r}_k\,...\,{\bf
p}_k\,...\,)\,\rightarrow\,F_{n-1}(...\,{\bf r}_{k-1}, {\bf
r}_{k+1}...\,\,{\bf
p}_{k-1}, {\bf p}_{k+1}...\,)\,G_m({\bf p}_k)\,\,\,,\\
F_n^{(eq)}(...\,{\bf
r}_k...\,)\,\rightarrow\,F_{n-1}^{(eq)}(...\,{\bf r}_{k-1}, {\bf
r}_{k+1}...\,)\,\,\,,
\end{array}
\end{equation}
when $\,{\bf r}_1\rightarrow \infty\,$ and $\,{\bf r}_k\rightarrow
\infty\,$, respectively.

It will be convenient, again following \cite{bog}, to collect all the
DF into one generating functional
\[
\mathcal{F}\{t,{\bf R},{\bf P},\psi\,|{\bf R}_0;\nu_{0}\}=F_0(t,{\bf
R},{\bf P}|{\bf R}_0;\nu_{0})+\sum_{n\,=1}^{\infty } \frac
{\nu_0^n}{n!}\int_1 ...\int_n F_n(t,{\bf R},{\bf r}^{(n)},{\bf
P},{\bf p}^{(n)}|{\bf R}_0;\nu_{0})\prod_{j\,=1}^n \psi({\bf
r}_j,{\bf p}_j)
\]
and correspondingly all the equations (\ref{f0})-(\ref{fn}) into
single equation for this functional:
\begin{equation}
\begin{array}{l}
\frac {\partial \mathcal{F}}{\partial t}\,+\,\frac {\bf
P}{M}\cdot\frac {\partial \mathcal{F}}{\partial {\bf
R}}\,=\,\mathcal{\widehat{L}}\left(\psi,\frac {\delta }{\delta
\psi}\right )\mathcal{F}\,\,\,\equiv\,-\int_1 \psi(x_1)\, \frac {{\bf
p}_1}{m}\cdot\frac {\partial }{\partial {\bf r}_1}\,\frac {\delta
\mathcal{F}}{\delta \psi(x_1)}\,\,\,+\,\,\label{fe}\\
+\,\int_1 [\,1+\psi(x_1)\,]\,\,\widehat{L}^b_1\,\,\frac {\delta
\mathcal{F}}{\delta \psi(x_1)}\,\,+\,\frac 12 \int_1\int_2
[\,1+\psi(x_1)\,]\,[\,1+\psi(x_2)\,]\, \,\widehat{L}^a_{12}\,\,\frac
{\delta^2 \mathcal{F}}{\delta \psi(x_1)\,\delta \psi(x_2)}\,\,
\end{array}
\end{equation}
For brevity here and below single argument\, $\,x_j\,$\, deputizes
for the pair $\,\{{\bf r}_j,{\bf p}_j\}\,$\,.

Equation (\ref{fe}) becomes direct analogue of equation (7.9) from
\cite{bog}\, if replace functional argument $\,u(x)\,$ by
$\,\nu_0\psi(x)\,$ and use obvious identity\, $\,\int_1\int_2
\widehat{L}^a_{12}\,...\,=0\,$\, \cite{bog}. Initial conditions
(\ref{ic}) take the form
\begin{equation}
\begin{array}{l}
\mathcal{F}\{0,\,{\bf R},{\bf P},\psi\,|{\bf R}_0;\nu_0\}\, =
\,\delta({\bf R}-{\bf R}_0)\,G_M({\bf P})\, \mathcal{F}^{(eq)}\{\phi
|{\bf R};\nu_0\}\,\,\,,\label{icf}
\end{array}
\end{equation}
where generating functional for the equilibrium DF and new argument
for this functional are introduced:
\[
\begin{array}{l}
\mathcal{F}^{(eq)}\{\phi |{\bf R};\nu_0\}\,= \,1\,
+\sum_{n\,=1}^{\infty }\frac {\nu_0^n}{n!}\int ...\int
F_n^{(eq)}({\bf r}_1...\,{\bf r}_n|{\bf R};\nu_0) \prod_{j\,=1}^n
\phi({\bf r}_j)\,d{\bf r}_j\,\,\,,\\
\phi({\bf r})\,\equiv\,\int \psi({\bf r},{\bf p})\,G_m({\bf
p})\,d{\bf p}\,\,
\end{array}
\]
From physical point of view, expression (\ref{icf}) represents
thermodynamical equilibrium, since fixation of start BP's position in
no way disturbs it. Nevertheless, (\ref{icf}) is not a fixed point of
equation (\ref{fe}). The latter will be achieved at
$\,t\rightarrow\infty\,$ only, when BP's position $\,{\bf R}(t)\,$
becomes completely uncertain. Hence, formal statistical equilibrium
is presented by functional $\,G_M({\bf P})\, \mathcal{F}^{(eq)}\{\phi
|{\bf R};\nu_0\}\,$, and we can write
\begin{equation}
\left[\,-\frac {\bf P}{M}\cdot\frac {\partial \mathcal{F}}{\partial
{\bf R}}\,+\,\mathcal{\widehat{L}}\left(\psi,\frac {\delta }{\delta
\psi}\right )\,\right ]\,G_M({\bf P})\, \mathcal{F}^{(eq)}\{\phi
|{\bf R};\nu_0\}\,=\,0\,\,\label{te}
\end{equation}
This equation easy transforms into another one in terms of
$\,\phi({\bf r})\,$ itself\,:
\begin{equation}
\left[\frac {\partial }{\partial {\bf r}}\,+\frac {\nabla \Phi_b({\bf
r}-{\bf R})}{T}\right]\frac {\delta \mathcal{F}^{(eq)}}{\delta
\phi({\bf r})}\,=\,\frac 1T \int [\,1+\phi({\bf
r}^{\prime})\,]\,\nabla \Phi_a({\bf r}^{\prime}-{\bf r})\, \frac
{\delta^2 \mathcal{F}^{(eq)}}{\delta \phi({\bf r})\,\delta \phi({\bf
r}^{\prime})}\,d{\bf r}^{\prime}\,\label{ter}
\end{equation}
This is analogue of equation (2.14) in \cite{bog}. As combined with
boundary conditions at infinity (\ref{unc}) it quite unambiguously
determines DF $\,F_n^{(eq)}\,$ and thus initial condition to equation
(\ref{fe}).

\section{Correlation functions and historical correlations}

Though DF $\,F_n^{(eq)}\,$ and thus $\,F_n(t=0,\,...\,)\,$ include
all equilibrium inter-particle correlations and at $\,t>0\,$ our
system remains in equilibrium, nevertheless at $\,t>0\,$ some excess
correlations do arise, so that
\[
\begin{array}{l}
F_n(t,{\bf R}, {\bf r}_1\,...\,{\bf r}_n,{\bf P},{\bf p}_1\,...\,{\bf
p}_n|{\bf R}_0;\nu_0)\,\neq\,F_0(t,{\bf R},{\bf P}|{\bf R}_0;\nu_0)
\,\,F_n^{(eq)}({\bf r}_1...\,{\bf r}_n|{\bf R};\nu_0)\prod_j G_m({\bf
p}_j)\,\,\,,
\end{array}
\]
as opposed to what one could think. To describe a difference between
left and right-hand side here, let us introduce ``correlation
functions'' (CF)\, $\,V_n(t,{\bf R},{\bf r}_1\,...\,{\bf r}_n,{\bf
P},{\bf p}_1\,...\,{\bf p}_n|{\bf R}_0;\nu_0)\,$\, and their
generating functional
\[
\mathcal{V}\{t,{\bf R},{\bf P},\psi\,|{\bf R}_0;\nu_{0}\}=V_0(t,{\bf
R},{\bf P}|{\bf R}_0;\nu_{0})+\sum_{n\,=1}^{\infty } \frac
{\nu_0^n}{n!}\int_1 ...\int_n V_n(t,{\bf R},{\bf r}^{(n)},{\bf
P},{\bf p}^{(n)}|{\bf R}_0;\nu_{0})\prod_{j\,=1}^n \psi({\bf
r}_j,{\bf p}_j)
\]
and define all them in the language of functionals be relation
\begin{equation}
\begin{array}{l}
\mathcal{F}\{t,{\bf R},{\bf P},\,\psi\,|{\bf R}_0;\nu_0\}\,\,=\,
\,\mathcal{F}^{(eq)}\{\phi\,|{\bf R};\nu_0\}\,\,\mathcal{V}\{t,{\bf
R},{\bf P},\,\psi\,|{\bf R}_0;\nu_0\}\,\,\,,\label{vf}
\end{array}
\end{equation}
where\, $\,\phi({\bf r})=\int \psi({\bf r},{\bf p})\,G_m({\bf
p})\,d{\bf p}\,$\, as in (\ref{icf}). The term ``correlation
functions'' is traditionally used in statistical mechanics for
various corrections to equilibrium or quasi-equilibrium distributions
(see e.g. \cite{uf,re,bal}). By this definition, in particular,
\[
F_0(t,{\bf R},{\bf P}| {\bf R}_0;\nu_0)=V_0(t,{\bf R},{\bf P}| {\bf
R}_0;\nu_0)\,\,\,,
\]
\begin{equation}
\begin{array}{l}
F_1(t,{\bf R},{\bf r}_1,{\bf P},{\bf p}_1|{\bf R}_0;\nu_0)=F_0(t,{\bf
R},{\bf P}|{\bf R}_0;\nu_0)\,F_1^{(eq)}({\bf r}_1|{\bf
R};\nu_0)\,G_m({\bf p}_1)+ V_1(t,{\bf R},{\bf r}_1,{\bf P},{\bf
p}_1|{\bf R}_0;\nu_0) \label{cf1}
\end{array}
\end{equation}
Initial conditions (\ref{ic}) transform to
\begin{equation}
\begin{array}{l}
V_n(t=0,...\,)\,=\,\delta_{n,\,0}\,\,\delta({\bf R}-{\bf R}_0)
\,\,\,,\,\,\,\,\,\,\mathcal{V}\{t=0,\,{\bf R},{\bf P},\,\psi\,|{\bf
R}_0;\nu_0\}\,=\,\delta({\bf R}-{\bf R}_0)\,\,\,,\label{icv}
\end{array}
\end{equation}
while the boundary conditions (\ref{unc}) prescribe
\begin{equation}
\begin{array}{l}
V_{n>\,0}(t\,,...\,{\bf r}_k...\,)\,\rightarrow
\,0\,\,\,\,\,\,\,\,\texttt{at}\,\,\,\,\,\,\,{\bf r}_k\,\rightarrow\,
\infty\,\,\label{bcv}
\end{array}
\end{equation}
It should be underlined that presence in $\,V_1\,$ of start position
of BP, $\,{\bf R}_0\,$, is principally important. As we already
mentioned, if $\,{\bf R}_0\,$ was unknown then then all the DF would
be invariable, that is all $\,V_{n>\,0}\,$ would turn to zero. This
remark highlights two peculiarities of excess correlations described
by $\,V_{n>\,0}\,$. First, they are essentially {\bf spatial}
correlations. Second, they connect current positions of atoms with
not merely current position of BP but its total previous
displacement, or {\bf path}, $\,{\bf R}-{\bf R}_0\,$, during time
interval $\,(0,t)\,$. By this reason we can name them ``historical
correlations''. Further we will see that their actual extent is very
closely related to statistics of the BP's path $\,{\bf R}-{\bf
R}_0\,$.

Substitution of (\ref{vf}) to equation (\ref{fe}) yields, with
accounting for identity (\ref{te}), equation for the generating
functional of correlation functions:
\begin{equation}
\frac {\partial \mathcal{V}}{\partial t}\,+\,\frac {\bf
P}{M}\cdot\frac {\partial \mathcal{V}}{\partial {\bf
R}}\,=\,\mathcal{L}\left(\psi,\frac {\delta }{\delta \psi}\right
)\,\mathcal{V}\,+\,\mathcal{L}^{\prime}\left(\nu_0,\psi,\frac {\delta
}{\delta \psi}\right )\,\mathcal{V}\,\,\,,\,\label{fev}
\end{equation}
where action of operator $\,\mathcal{L}\,$ is described by expression
(\ref{fe}) while action of new operator $\,\mathcal{L}^{\prime}\,$ by
\begin{equation}
\mathcal{L}^{\prime}\left(\nu_0,\psi,\frac {\delta }{\delta
\psi}\right )\mathcal{V}\,=\,\left\{\int [\,1\,+\phi({\bf
r})\,]\,\,\nabla \Phi_b({\bf R}-{\bf r})\,\frac {\delta \ln
\mathcal{F}^{(eq)}(\phi |{\bf R};\nu_0)}{\delta \phi({\bf
r})}\,\,d{\bf r}\right\}\left(\frac {{\bf P}}{MT}+\frac {\partial
}{\partial {\bf P}}\right )\,\mathcal{V}\,\,+\,\,\,\label{lopc}
\end{equation}
\[
\,\,\,\,\,\,\,\,\,\,\,\,\,\,\,\,\,\,\,\,\,\,\,\,\,\,\,\,\,
\,\,\,\,\,\,\,\,\,\,\,\,\,\,\,\,\,\,\,+\,\,\int_1\int_2
[\,1+\psi(x_1)\,]\,[\,1+\psi(x_2)\,]\,\,\widehat{L}^a_{12}\,\,\frac
{\delta \ln \mathcal{F}^{(eq)}(\phi |{\bf R};\nu_0)}{\delta \phi({\bf
r}_2)}\,\,G_m({\bf p}_2)\,\frac {\delta \,\mathcal{V}}{\delta
\psi(x_1)}\,
\]
So complicated expression is payment for simple initial condition
(\ref{icv}). After that, variational derivatives of (\ref{fev}) at
$\,\psi=0\,$ give BBGKY equation in terms of the CF. The first of
them looks quite similar to (\ref{f0}):
\begin{equation}
\frac {\partial V_0}{\partial t}\,=\,-\frac {{\bf P}}{M}\cdot\frac
{\partial V_0}{\partial {\bf R}}\, +\nu_0\int_1
\widehat{L}_{1}^{b}\,V_1\,\,\label{v0}
\end{equation}
Since $\,V_0\equiv F_0\,$, comparison of (\ref{v0}) with (\ref{f0})
shows that quasi-equilibrium part of the pair DF $\,F_1\,$, i.e.
first right-hand term  in (\ref{cf1}), does not contribute to the
collision term in (\ref{f0}), and collisions of BP with atoms are
wholly represented by pair CF $\,V_1\,$. The second BBGKY equation
becomes
\begin{equation}
\begin{array}{l}
\frac {\partial V_1}{\partial
t}\,=\,\widehat{L}_{1}\,V_1\,+\,\nu_0\int_2
\widehat{L}_2^{b}\,V_2\,+\,\nu_0
\int_2\widehat{L}_{1\,2}^{a}\,[\,V_2\,+\,F_1^{(eq)}({\bf
1})\,G_m({\bf 1})\,V_1({\bf 2})+F_1^{(eq)}({\bf 2})\,G_m({\bf
2})\,V_1({\bf 1})\,]\,\,+\label{v1}\\ +\,\,G_m({\bf 1})\,\left
\{\,\nabla \Phi_b({\bf R}-{\bf r}_1)\,F_1^{(eq)}({\bf
1})\,+\,\nu_0\int \nabla\Phi_b({\bf R}-{\bf r})\,C_2^{(eq)}({\bf
r},{\bf r}_1|{\bf R};\nu_0)\,d{\bf r}\right \} \left(\frac {{\bf
P}}{MT}+\frac {\partial }{\partial {\bf P}}\right )V_0\,
\end{array}
\end{equation}
with\,\, $\,F_1^{(eq)}({\bf k})\equiv F_1^{(eq)}({\bf r}_k|{\bf
R};\nu_0)\,$,\, $\,V_1({\bf k})\equiv V_1(t,{\bf R},{\bf r}_k,{\bf
P},{\bf p}_k|{\bf R}_0;\nu_0)\,$\, and $\,G_m({\bf k})\equiv G_m({\bf
p}_k)\,$.\, The function
\[
\,\,C_2^{(eq)}({\bf r}_1,{\bf r}_2|{\bf R};\nu_0)\,\equiv
\,F_2^{(eq)}({\bf r}_1,{\bf r}_2|{\bf R};\nu_0)-F_1^{(eq)}({\bf
r}_1|{\bf R};\nu_0)\,F_1^{(eq)}({\bf r}_2|{\bf R};\nu_0)\,
\]
is pair CF of the fluid under given BP's position.

\section{Equilibrium generating functional and its invariant
transformations}

Next, consider more carefully the equilibrium generating functional
$\,\mathcal{F}^{(eq)}\,$. It is of interest for us here since
operator (\ref{lopc}) includes its logarithmic variational
derivative. Let us write the latter as
\begin{equation}
\begin{array}{l}
\frac {\delta \ln \mathcal{F}^{(eq)}\{\phi |{\bf R};\,\nu_0\}}{\delta
\phi({\bf r})}\,=\,\nu_0\,\mathcal{C}\{{\bf r},\phi\,|{\bf
R};\,\nu_0\}\,=\,\label{c}\\
\,\,\,\,\,\,\,\,\,\,\,\,\,\,\,\,\,\,\,\,\,\,\,\,\,\,\,\,\,\,=\,\nu_0
\left[\,1+C_1^{(eq)}({\bf r}|{\bf R};\nu_0)+\sum_{n\,=\,1}^{\infty
}\frac {\nu_0^n}{n!}\int ...\int C_{n+1}^{(eq)}({\bf r},{\bf
r}_1...\,{\bf r}_n|\,{\bf R};\nu_0) \prod_{j\,=1}^n \phi({\bf
r}_j)\,d{\bf r}_j\,\right ]\,\,\,,
\end{array}
\end{equation}
where\, $\,C_1^{(eq)}({\bf r}|{\bf R};\nu_0)\,= \,F_1^{(eq)}({\bf
r}|{\bf R};\nu_0)-1\,$,\, pair correlation function
$\,C_2^{(eq)}({\bf r},{\bf r}_1|{\bf R};\nu_0)\,$\, was presented
just above, third-order correlation function or, to be precise,
cumulant function is
\[
\begin{array}{l}
C_3^{(eq)}({\bf r},{\bf r}_1,{\bf r}_2)\,=\,F^{(eq)}_3({\bf r},{\bf
r}_1,{\bf r}_2)\,+\,2\,F^{(eq)}_1({\bf r})\,F^{(eq)}_1({\bf
r}_1)\,F^{(eq)}_1({\bf r}_2)\,-\\
\,\,\,\,\,\,\,\,\,\,\,\,\,\,\,\,\,\,\,\,\,\,\,\,\,\,\,\,\,
\,\,\,\,\,\,\,\,\,\,\,\, \,\,\,\,\,\,\,\,\,\,\,\,\,\,
-\,F^{(eq)}_1({\bf r})\,F^{(eq)}_2({\bf r}_1,{\bf
r}_2)\,-\,F^{(eq)}_1({\bf r}_1)\,F^{(eq)}_2({\bf r},{\bf
r}_2)\,-\,F^{(eq)}_1({\bf r}_2)\,F^{(eq)}_2({\bf r},{\bf r}_1)\,\,\,,
\end{array}
\]
for brevity omitting arguments $\,{\bf R}\,$ and $\,\nu_0\,$, and so
on. Then equation (\ref{ter}) transforms to
\begin{equation}
\begin{array}{l}
\left[\frac {\partial }{\partial {\bf r}}\,+\frac {\nabla \Phi_b({\bf
r}-{\bf R})}{T}\right]\mathcal{C}^{(eq)}\{{\bf r},\phi\,|{\bf
R};\,\nu_0\}\,=\,\frac 1T \int [\,1+\phi({\bf r}^{\prime})\,]\,\nabla
\Phi_a({\bf r}^{\prime}-{\bf r})\,\frac {\delta
\mathcal{C}^{(eq)}\{{\bf r},\phi\,|{\bf R};\,\nu_0\}}{\delta
\phi({\bf r}^{\,\prime})}\,\,d{\bf r}^{\prime}\,+\\
\,\,\,\,\,\,\,\,\,\,\,\,\,\,\,\,\,\,\,\,\,\,\,\,\,\,
\,\,\,\,\,\,\,\,\,\,\,\,+\,\,\mathcal{C}^{(eq)}\{{\bf r},\phi\,|{\bf
R};\,\nu_0\}\,\,\frac {\nu_{\,0}}{T}\int [\,1+\phi({\bf
r}^{\prime})\,]\,\nabla \Phi_a({\bf r}^{\prime}-{\bf
r})\,\mathcal{C}^{(eq)}\{{\bf r}^{\prime},\phi\,|{\bf
R};\,\nu_0\}\,d{\bf r}^{\,\prime}\,\,\label{ter1}
\end{array}
\end{equation}
This nonlinear equation can not be solved in a closed form but it
will help us to perceive important features of functional
$\,\mathcal{C}^{(eq)}\,$.

With this purpose, first, notice that due to the DF's asymptotic at
infinity (\ref{unc}) at any reasonable potentials of atom-atom and
atom-BP interactions, similar to (\ref{bcv}),
\begin{equation}
\begin{array}{l}
C_{n+1}^{(eq)}({\bf r},{\bf r}_1...\,{\bf r}_n|{\bf
R};\nu_0)\,\rightarrow\,0\,\,\,\,\,\,\,\,\,\texttt{if}\,\,\,\,\,\,\,\,{\bf
r}_j-{\bf r}\rightarrow\infty\,\,\,,\label{cbc}\\
C_1^{(eq)}({\bf r}_1|{\bf
R};\nu_0)\,\rightarrow\,0\,\,\,\,\,\,\,\,\texttt{if}\,\,\,\,\,\,\,\,\,{\bf
R}-{\bf r}_1\rightarrow\infty\,\,\,,\\
C_{n>\,1}^{(eq)}({\bf r}_1...\,{\bf r}_n|{\bf
R};\nu_0)\,\rightarrow\,C_{n>\,1}^{(eq)}({\bf r}_1...\,{\bf
r}_n|\nu_0)\,\,\,\,\,\,\,\,\texttt{if}\,\,\,\,\,\,\,\,\,{\bf R}-{\bf
r}_j\rightarrow\infty\,\,\,,
\end{array}
\end{equation}
where cumulant functions $\,C_n^{(eq)}({\bf r}_1...\,{\bf
r}_n|\nu_0)\,$ describe fluid without BP and satisfy
\[
\begin{array}{l}
C_{n+1}^{(eq)}({\bf r},{\bf r}_1...\,{\bf
r}_n|\,\nu_0)\,\rightarrow\,0\,\,\,\,\,\,\,\,\,\texttt{if}\,\,\,\,\,\,\,\,{\bf
r}_j-{\bf r}\rightarrow\infty\,
\end{array}
\]
Moreover, any of the cumulant functions $\,C_{n>\,1}^{(eq)}({\bf
r}_1...\,{\bf r}_n|{\bf R};\nu_0)\,$ tend to zero at $\,{\bf
r}_j\rightarrow\infty\,$ in so fast way that all they are integrable
in respect to $\,{\bf r}_1...\,{\bf r}_n\,$.

Consequently, functional $\,\mathcal{C}^{(eq)}\,$ keeps a sense when
$\,\phi({\bf r})\,$ turns into nonzero constant, $\,\phi({\bf
r})\rightarrow\sigma =\,$\,const\,, as well as when one replaces
$\,\phi({\bf r})\,$ by $\,\sigma +\phi({\bf r})\,$ with $\,\sigma
=\,$\,const\,. Thus we can introduce function
\begin{equation}
\begin{array}{l}
C(\sigma ,\nu_0)=\lim_{{\bf R}-\,{\bf r}\,\rightarrow\infty}
\,\mathcal{C}^{(eq)}\{{\bf r},\sigma|{\bf
R};\nu_0\}=1+\sum_{n\,=\,1}^{\infty }\frac {\nu_0^n\sigma^n}{n!}\int
...\int C_{n+1}^{(eq)}({\bf r},{\bf r}_1...\,{\bf
r}_n|\nu_0)\,\,d{\bf r}_1...\,d{\bf r}_n\,\,\,,\label{ccc}
\end{array}
\end{equation}
where, clearly, integrals factually do not depend on $\,\,{\bf r}\,$.
Besides, introduce new functional
\begin{equation}
\begin{array}{l}
\mathcal{C}^{(eq)}_{\sigma}\{{\bf r},\phi\,|{\bf R}\}\,=\,\frac
{\mathcal{C}^{(eq)}\{{\bf r},\sigma +\phi\,|{\bf
R};\,\nu_0\}}{\lim_{\,{\bf R}-\,{\bf r}\,\rightarrow\infty}
\,\mathcal{C}^{(eq)}\{{\bf r},\sigma|{\bf R};\nu_0\}} \,=\,\frac
{\mathcal{C}^{(eq)}\{{\bf r},\sigma +\phi\,|{\bf
R};\,\nu_0\}}{C(\sigma ,\nu_0)}\,\label{nc}
\end{array}
\end{equation}
Under change $\,\phi({\bf r})\rightarrow\sigma +\phi({\bf r})\,$
equation (\ref{ter1}) evidently turns to equation for
$\,\mathcal{C}^{(eq)}_{\sigma}\{{\bf r},\phi\,|{\bf R}\}\,$\, as
follows,
\begin{eqnarray}
\begin{array}{c}
\,\,\,\left[\frac {\partial }{\partial {\bf r}}\,+\frac {\nabla
\Phi_b({\bf r}-{\bf R})}{T}\right]\mathcal{C}^{(eq)}_{\sigma}\{{\bf
r},\phi\,|{\bf R}\}\,=\,\frac 1T \int \left[\,1+\frac {\phi({\bf
r}^{\prime})}{1+\sigma}\,\right]\,\nabla \Phi_a({\bf r}^{\prime}-{\bf
r})\,\frac {\delta\, \mathcal{C}^{(eq)}_{\sigma}\{{\bf r},\phi\,|{\bf
R}\}}{\delta\,[\, \phi({\bf r}^{\,\prime})/(1+\sigma )\,]}\,\,d{\bf
r}^{\prime}\,+\label{ter2}\\
+\,\,\mathcal{C}^{(eq)}_{\sigma}\{{\bf r},\phi\,|{\bf R}\}\,\,\frac
{\nu_{\,0}\,C(\sigma ,\nu_0)\,(1+\sigma)}{T}\int \left[\,1+\frac
{\phi({\bf r}^{\prime})}{1+\sigma}\,\right]\,\nabla \Phi_a({\bf
r}^{\prime}-{\bf r})\,\,\mathcal{C}^{(eq)}_{\sigma}\{{\bf
r}^{\,\prime},\phi\,|{\bf R}\}\,\,d{\bf r}^{\,\prime}\,\,
\end{array}
\end{eqnarray}

Second, a simple analysis of structure of equation (\ref{ter1}) shows
that it uniquely determines its solution in the form of expansion
(\ref{c}) if coefficients of the expansion satisfy conditions
(\ref{cbc}).

Third, coefficients of similar series expansion of the functional
(\ref{nc}) satisfy absolutely similar conditions, because, in
consequence of (\ref{cbc}),
\begin{equation}
\begin{array}{l}
\int ...\int C_{n+k+1}^{(eq)}({\bf r},{\bf r}_1...\,{\bf r}_n,\,{\bf
\rho}_1...\,{\bf \rho}_k|{\bf R};\nu_0)\,d{\bf \rho}_1...\,d{\bf
\rho}_k\,\rightarrow\,0\,\,\,\,\,\,\,\,\,\texttt{if}\,\,\,\,\,\,\,\,{\bf
r}_j-{\bf r}\rightarrow\infty\,\,\,,\label{cbc1}\\
\mathcal{C}^{(eq)}_{\sigma}\{{\bf r},\phi =0\,|{\bf
R}\}\,\rightarrow\,1\,\,\,\,\,\,\,\,\texttt{if}\,\,\,\,\,\,\,\,\,{\bf
R}-{\bf r}\rightarrow\infty\,\,
\end{array}
\end{equation}

Fourth, the only formal difference of equation (\ref{ter2}) from
(\ref{ter1}) is mere change of its two independent variables,
concretely,\, $\,\phi({\bf r})\rightarrow \phi({\bf r})/(1+\sigma
)\,$\, and\, $\,\nu_{\,0}\rightarrow \nu(\sigma ,\nu_0)\,$,\, where
\begin{equation}
\begin{array}{l}
\nu (\sigma ,\nu_0)\,=\,\nu_{\,0}\,C(\sigma
,\nu_0)\,(1+\sigma)\,\,\label{nd}
\end{array}
\end{equation}

These observations taken together mean that a proper solution to
equation (\ref{ter2}),\, $\,\mathcal{C}^{(eq)}_{\sigma}\{{\bf
r},\phi\,|{\bf R}\}\,$,\,  is nothing but $\,\mathcal{C}^{(eq)}\{{\bf
r},\phi/(1+\sigma )\,|{\bf R};\,\nu (\sigma ,\nu_0) \}\,$,\, that is,
in view of (\ref{nc}) and (\ref{nd}), for any $\,\sigma
=\,$\,const\,\, identity
\begin{equation}
\begin{array}{l}
\nu_0\,\mathcal{C}^{(eq)}\{{\bf r},\sigma +\phi\,|\,{\bf
R};\,\nu_0\}\,=\,\frac {\nu (\sigma
,\nu_0)}{1+\sigma}\,\,\mathcal{C}^{(eq)}\left\{{\bf r},\frac
{\phi}{1+\sigma}\,|\,{\bf R};\,\nu (\sigma
,\nu_0)\right\}\,\,\,\label{eqid}
\end{array}
\end{equation}
takes place. In essence two latter formulas define a group of such
transformations of equilibrium generating functional
$\,\mathcal{C}^{(eq)}\{{\bf r},\phi\,|\,{\bf R};\,\nu_0\}\,$ which do
not change its value:
\begin{equation}
\begin{array}{l}
\widehat{\mathcal{T}}(\sigma)\,\mathcal{C}^{(eq)}\{{\bf
r},\phi\,|\,{\bf R};\,\nu_0\}\,\equiv\,\frac {\nu (\sigma ,\nu_0
)}{(1+\sigma)\,\nu_0}\,\,\mathcal{C}^{(eq)}\left\{{\bf r},\frac
{1+\phi}{1+\sigma}-1\,|\,{\bf R};\,\nu (\sigma ,\nu_0)\right\}\,=
\,\mathcal{C}^{(eq)}\{{\bf r},\phi\,|\,{\bf
R};\,\nu_0\}\,\,\,,\label{gr}\\
\widehat{\mathcal{T}}(\sigma_2)\,\widehat{\mathcal{T}}(\sigma_1)\,=
\,\widehat{\mathcal{T}}(\sigma_1 +\sigma_2
+\sigma_1\sigma_2)\,\,\,,\\
\nu (\sigma_2\,,\,\nu (\sigma_1 ,\nu_0))\,=\,\nu (\sigma_1 +\sigma_2
+\sigma_1\sigma_2 \,,\nu_0)
\end{array}
\end{equation}
At that, obviously, the group parameter should satisfy $\,\sigma
>-1\,$, and similar restriction should be imposed upon $\,\phi({\bf
r})\,$. But this restriction disappears if write $\,\sigma
=\exp{(a)}-1\,$,\, so that in terms of new parameter
$\,\widehat{\mathcal{T}}(a_2)\,\widehat{\mathcal{T}}(a_1)\,=
\,\widehat{\mathcal{T}}(a_1 +a_2)\,$.

Infinitesimal form of functional identities (\ref{eqid}) or
(\ref{gr}) looks better in terms of particular cumulants:
\begin{eqnarray}
\begin{array}{l}
\left\{n\,\varkappa(\nu)\,+\,[\,1+\varkappa(\nu)\,] \,\nu\,\frac
{\partial}{\partial \nu}\,\right\}[\,C_{n}^{(eq)}({\bf r}_1...\,{\bf
r}_n|{\bf R};\nu)+\delta_{\,n,\,1}]\,=\,\nu\int C_{n+1}^{(eq)}({\bf
r}_1...\,{\bf r}_{n},{\bf r}^{\prime}|{\bf R};\nu)\,d{\bf
r}^{\prime}\,\,\,,\label{eqinf}\\
\varkappa(\nu)\,\equiv \,\left[\frac {\partial C(\sigma
,\nu)}{\partial \sigma}\right]_{\sigma =0}\,=\,\nu\int
C_2^{(eq)}({\bf r},0|\,\nu)\,d{\bf r}\,
\end{array}
\end{eqnarray}
Recall that\,
\[
1+\varkappa(\nu)\,=\,T\,\left(\frac {\partial \nu}{\partial
P}\right)_T\,\,\,,
\]
with $\,P\,$ being pressure, is isothermal compressibility of a fluid
\cite{re,bal,ll1}.

\section{Invariance group of the historical correlations functional
and virial relations}

Now, turn our attention to evolution equation (\ref{fev}) for the
generating functional of correlation functions,
$\,\mathcal{V}\{t,{\bf R},{\bf P},\psi\,|{\bf R}_0;\nu_{0}\}\,$, and
consider its principal properties implied by structure of operator
$\,\mathcal{L}+\mathcal{L}^{\,\prime}\,$, above mentioned properties
of factor\, $\,\delta \ln \mathcal{F}^{(eq)}\{\phi |{\bf
R};\,\nu_0\}/\delta \phi({\bf
r})\,=\,\nu_{0}\,\mathcal{C}^{(eq)}\{{\bf r},\phi |{\bf
R};\,\nu_0\}\,$\, which enters\, $\,\mathcal{L}^{\,\prime}\,$,\,
initial condition (\ref{icv}) and, of course, asymptotic boundary
conditions (\ref{bcv}).

What is important, the initial condition for $\,\mathcal{V}\,$ does
not include independent variables $\,\psi =\psi({\bf r},{\bf p})\,$
and $\,\nu_{0}\,$ at all. Therefore all effects of their
transformations are purely determined by properties of
$\,\mathcal{L}+ \mathcal{L}^{\,\prime}\,$ itself under boundary
conditions (\ref{bcv}). One role of the latter is as just above: they
allow to extend functional $\,\mathcal{V}\{t,{\bf R},{\bf
P},\psi\,|{\bf R}_0;\nu_{0}\}\,$ to argument $\,\sigma +\psi({\bf
r},{\bf p})\,$ with $\,\sigma =\,$\,const\, in place of  $\,\psi({\bf
r},{\bf p})\,$. Another role of conditions (\ref{bcv}) is that due to
them, and according to definition (\ref{fe}) of operator
$\,\widehat{\mathcal{L}}\,$,\, variable $\,\psi(x_1)\,$ in the first
term of\, $\,\mathcal{\widehat{L}}\,\mathcal{V}\,$\, can be shifted
by arbitrary constant:
\[
\int_1 \psi(x_1)\, \frac {{\bf p}_1}{m}\cdot\frac {\partial
}{\partial {\bf r}_1}\,\frac {\delta \mathcal{V}}{\delta
\psi(x_1)}\,\,=\,\int_1 [\,\sigma +\psi(x_1)\,]\, \frac {{\bf
p}_1}{m}\cdot\frac {\partial }{\partial {\bf r}_1}\,\frac {\delta
\mathcal{V}}{\delta \psi(x_1)}\,\,\,,
\]
where $\,\sigma =\,$\,const\,, e.g. $\,\sigma =1\,$. This is
principal difference of expression
$\,\mathcal{\widehat{L}}\,\mathcal{V}\,$ from
$\,\mathcal{\widehat{L}}\,\mathcal{F}\,$. As the consequence, taking
in mind action of $\,\widehat{\mathcal{L}}\,$ onto $\,\mathcal{V}\,$,
one can write
\begin{equation}
\mathcal{\widehat{L}}\left(\sigma +\psi\,,\frac {\delta }{\delta
\psi}\right )\,=\,\mathcal{\widehat{L}}\left(\frac
{\psi}{1+\sigma}\,,\frac {\delta }{\delta [\,\psi/(1+\sigma
)\,]}\right )\,\,\label{lt}
\end{equation}
Visual study of operator $\,\mathcal{\widehat{L}}^{\,\prime}\,$
defined by (\ref{lopc}), with taking into account identity
(\ref{eqid}) for\, $\,\nu_{0}\,\mathcal{C}^{(eq)}\{{\bf r},\phi |{\bf
R};\,\nu_0\}\,$\,, shows that $\,\mathcal{\widehat{L}}^{\,\prime}\,$
possesses the same invariance property, if transformation of
$\,\psi({\bf r},{\bf p})\,$ is accompanied by transformation of the
density argument $\,\nu_0\,$ in conformity with (\ref{nd}) and
(\ref{ccc}):
\begin{equation}
\mathcal{\widehat{L}}^{\,\prime}\left(\nu_0\,,\,\sigma +\psi\,,\frac
{\delta }{\delta \psi}\right
)\,=\,\mathcal{\widehat{L}}^{\,\prime}\left(\,\nu(\sigma,\nu_0)\,,\,\frac
{\psi}{1+\sigma}\,,\frac {\delta }{\delta [\,\psi/(1+\sigma
)\,]}\right )\,\,\label{lpt}
\end{equation}

Because of (\ref{lt}) and (\ref{lpt}) equation (\ref{fev}) under
initial condition (\ref{icv}) implies similar invariance property for
its solution:
\begin{equation}
\begin{array}{l}
\mathcal{V}\{t,{\bf R},{\bf P},\,\sigma +\psi\,|\,{\bf
R}_0;\,\nu_0\}\,=\,\mathcal{V}\left\{t,{\bf R},{\bf P},\,\frac
{\psi}{1+\sigma}\,|\,{\bf R}_0;\,\nu (\sigma
,\nu_0)\right\}\,\,\,\label{id}
\end{array}
\end{equation}
with arbitrary $\,\sigma ({\bf r}) =\,$\,const\,. In other words, the
generating functional $\,\mathcal{V}\,$ of time-dependent
``historical correlations'' also is invariant under above mentioned
group of transformations:
\begin{equation}
\begin{array}{l}
\widehat{\mathcal{T}}(\sigma)\,\mathcal{V}\{t,{\bf R},{\bf
P},\,\psi\,|\,{\bf R}_0;\,\nu_0\}\,\equiv\,\mathcal{V}\{t,{\bf
R},{\bf P},\,\frac {1+\psi}{1+\sigma}-1\,|\,{\bf R}_0;\,\nu (\sigma
,\nu_0)\}\,=\,\mathcal{V}\{t,{\bf R},{\bf P},\,\psi\,|\,{\bf
R}_0;\,\nu_0\}\,\,\,,\,\label{grv}
\end{array}
\end{equation}
where left identity, as combined with (\ref{nd}) and (\ref{ccc}),
defines action of the group onto $\,\mathcal{V}\,$.

This is main formal result of the present paper. It would be
difficult to obtain this result from a chain of equations for
particular DF, (\ref{f0})-(\ref{fn}) or all the more (\ref{v0}),
(\ref{v1}), etc. Thus, perhaps, for the first time a real profit of
Bogolyubov's generating functional for non-equilibrium DF has been
demonstrated.

Infinitesimal form of (\ref{grv}) reads
\begin{eqnarray}
\begin{array}{l}
\left\{n\,\varkappa(\nu)\,+\,[\,1+\varkappa(\nu)\,] \,\nu\,\frac
{\partial}{\partial \nu}\,\right\}\,V_n(t,{\bf R},{\bf r}^{(n)},{\bf
P},{\bf p}^{(n)}|{\bf R}_0;\,\nu)\,=\,\\ \,\,\,\,\,\,\,\,\,\,\,\,
\,\,\,\,\,\,\,\,\,\,\,\,\,\,\,\,\,\,\,\,\,\,\,\,\,\,\,\,\,\,\,\,
\,\,\,\,\,\,\,\,\,\,\,\,\,\,\,\,\,\,\,\,\,\,\,\,\,\,\,\,\,\,\,\,
\,\,\,\,\,\,\,\,\,\,\,\,\,\,\,\,\,\,\,\,=\,\nu\int_{n+1}
V_{n+1}(t,{\bf R},{\bf r}^{(n+1)},{\bf P},{\bf p}^{(n+1)}|{\bf
R}_0;\,\nu) \,\,\,,\label{inf}
\end{array}
\end{eqnarray}
thus at $\,n>0\,$ connecting derivative of any BP-atoms correlation
in respect to density of atoms and next-order correlation integrated
over variables of additional outer atom. At $\,n=0\,$ we connect
derivative of the very BP's probability distribution and integrated
pair historical correlation:
\begin{eqnarray}
\begin{array}{l}
[\,1+\varkappa(\nu)\,] \,\frac {\partial V_0(t,{\bf R},{\bf P}|{\bf
R}_0;\,\nu)}{\partial \nu}\,=\,\int_{1} V_1(t,{\bf R},{\bf r}_1,{\bf
P},{\bf p}_1|{\bf R}_0;\,\nu)\,\,\,\label{inf0}
\end{array}
\end{eqnarray}
with $\,\varkappa(\nu)\,$ defined in (\ref{eqinf}).

From the other hand, if considering $\,\psi({\bf r},{\bf p})\,$ as
infinitesimal parameter instead of $\,\sigma\,$ and performing in
(\ref{id}) $\,k$-order variational differentiation over $\,\psi\,$ at
point $\,\psi=0\,$, one comes to rather different type of relations.
At $\,k>0\,$,
\begin{eqnarray}
\begin{array}{l}
\left[\frac {\nu (\sigma
,\nu_0)}{(1+\sigma)\nu_0}\right]^k\,V_k(t,{\bf R},{\bf r}^{(k)},{\bf
P},{\bf p}^{(k)}|{\bf R}_0;\,\nu (\sigma ,\nu_0))\,=\,\\
\,\,\,\,\,\,\,\, =\,V_k(t,{\bf R},{\bf r}^{(k)},{\bf P},{\bf
p}^{(k)}|{\bf R}_0;\,\nu_0)\,+\sum_{n\,=\,1}^\infty \frac
{\nu_0^n\sigma^n}{n!}\int_{k+1} ...\int_{k+n} V_{k+n}(t,{\bf R},{\bf
r}^{(k+n)},{\bf P},{\bf p}^{(k+n)}|{\bf
R}_0;\,\nu_{0})\,\,\label{vexpn}
\end{array}
\end{eqnarray}
For $\,k=0\,$, i.e. merely under putting on $\,\psi=0\,$, identity
(\ref{id}) yields
\begin{eqnarray}
\begin{array}{c}
V_0(t,{\bf R},{\bf P}|{\bf R}_0;\,\nu (\sigma ,\nu_0))
\,=\,V_0(t,{\bf R},{\bf P}|{\bf R}_0;\,\nu_{0})+\sum_{n\,=\,1}^\infty
\frac {\nu_0^n\sigma^n}{n!}\,\int_1 ...\int_n V_n(t,{\bf R},{\bf
r}^{(n)},{\bf P},{\bf p}^{(n)}|{\bf R}_0;\,\nu_{0})\,\,\label{vexp}
\end{array}
\end{eqnarray}
Of course, all terms in this formula actually depend on the BP's path
$\,\Delta{\bf R}={\bf R}-{\bf R}_0\,$ only.

Formulas (\ref{vexp}) and (\ref{vexpn}), by their sense, can be
interpreted as a sort of virial expansion of the BP's probability
distribution and BP's correlations with fluid, with those peculiarity
that in opposite to usual virial expansions of thermodynamic
quantities \cite{ll1} or kinetic coefficients \cite{ll2} our ones are
expansions over relative change of density, $\,\nu/\nu_0-1\,$, rather
than its absolute value. Then corresponding differential forms
(\ref{inf0}) and (\ref{inf}) can be named ``virial relations''.

\section{Some principal consequences of the virial relations}

Although invariance property (\ref{id}) of the functional
$\,\mathcal{V}\,$ does not substitute solving the equation
(\ref{fev}) of its evolution, corresponding virial relations bring
important information about structure of correlations functions and,
eventually, character of evolution of the BP's probability
distribution
\[
V_0(t,\Delta{\bf R},{\bf P}|\,\nu)\,\equiv\,V_0(t,{\bf R},{\bf
P}|{\bf R}_0;\,\nu)
\]
For example, let us consider relation (\ref{inf0}) supposing for
simplicity that fluid density $\,\nu\,$ in (\ref{inf0}) corresponds
to a dilute gas: $\,4\pi r_a^3\nu/3\ll 1\,$, $\,4\pi r_b^3\nu/3\ll
1\,$ (characteristic radii of interactions $\,r_a\,$ and $\,r_b\,$
were introduced above). Then $\,\varkappa(\nu)\ll 1\,$ and, as it is
known (see e.g. \cite{ll2}), mean free path of our molecular-size BP
and mean free path of atoms both are inversely proportional to gas
density:\, $\,\Lambda_b\sim (\pi r_b^2\nu)^{-1}\,$,\,
$\,\Lambda_a\sim (\pi r_a^2\nu)^{-1}\,$,\, together with diffusivity
of BP,\, $\,D_b=D_b(\nu )\sim v_b\Lambda_b\propto \nu^{-1}\,$,\, and
diffusivity of atoms, $\,D_a(\nu )\sim v_a\Lambda_a\propto
\nu^{-1}\,$,\, where\, $\,v_b=\sqrt{T/M}\,$ and $\,v_a= \sqrt{T/m}\,$
are thermal velocities. Besides, we take in mind large enough
temporal and spatial scales:\, $\,|\Delta {\bf R}|\gg \Lambda_b\,$
and $\,t\gg \tau_b \,$\, with $\,\tau_b \sim\Lambda_b /v_b\,$ being
mean free-flight time of BP.

First, it is useful to compare relation (\ref{inf0}), along with
(\ref{inf}), and relations (\ref{eqinf}). In the latter, in general,
$\,\max |C_{n}^{(eq)}|\sim 1\,$ while characteristic scale of
equilibrium inter-atom correlations equals to\, $\,r_a\,$\,.
Therefore right-hand sides of (\ref{eqinf}) can be roughly but surely
estimated as
\[
\begin{array}{l}
\int C_{n+1}^{(eq)}({\bf r}_1...\,{\bf r}_{n},{\bf r}^{\prime}|{\bf
R};\nu)\,d{\bf r}^{\prime}\,\sim\,r_a^3\,C_{n}^{(eq)}({\bf
r}_1...\,{\bf r}_n|{\bf R};\nu)\,\,
\end{array}
\]
On the left, consequently,
\[
\begin{array}{l}
\nu\,\frac {\partial C_{n}^{(eq)}({\bf r}_1...\,{\bf r}_n|{\bf
R};\nu)}{\partial \nu}\,\sim\,\,\nu \,r_a^3\,\,C_{n}^{(eq)}({\bf
r}_1...\,{\bf r}_n|{\bf R};\nu)\,\ll\,C_{n}^{(eq)}({\bf r}_1...\,{\bf
r}_n|{\bf R};\nu)\,\,\,,
\end{array}
\]
in particular,\, $\,\varkappa(\nu)\ll 1\,$\,. In opposite, in
(\ref{inf})-(\ref{inf0}) we can at first roughly but surely estimate
left sides:
\[
\begin{array}{l}
\nu\,\frac {\partial V_n(t,{\bf R},{\bf r}^{(n)},{\bf P},{\bf
p}^{(n)}|{\bf R}_0;\,\nu)}{\partial \nu}\,\sim\,V_n(t,{\bf R},{\bf
r}^{(n)},{\bf P},{\bf p}^{(n)}|{\bf R}_0;\,\nu)\,\,
\end{array}
\]
Indeed, a relatively small variation of gas density results in
similar relative variation of BP's diffusivity and therefore in
comparable relative change of probability of its displacement, as
well as connected probabilities and correlations. For instance, if
the hypothetical Gaussian asymptotic \cite{re,bal,dor} of BP's
distribution at $\,t\gg \tau_b \,$ and $\,|\Delta {\bf R}|\sim
\sqrt{D_bt}\gg \Lambda_b\,$\, realized,
\begin{equation}
\begin{array}{c}
V_0(t,\Delta{\bf R},{\bf P}|\,\nu)\,\rightarrow\,V_G(t,\Delta{\bf
R},{\bf P}|\,\nu)\,\equiv\,[4\pi D_b(\nu)t\,]^{-\,3/2}\,\exp{[-\Delta
{\bf R}^2/4D_b(\nu)t\,]}\,\,G_M({\bf P})\,\,\,, \label{vg}
\end{array}
\end{equation}
then one would have, evidently,
\begin{equation}
\begin{array}{l}
\nu\,\frac {\partial V_0(t,\Delta{\bf R},{\bf P}|\,\nu)}{\partial
\nu}\,\approx\,V_0(t,\Delta{\bf R},{\bf P}|\,\nu)\left [\,\frac 32
-\frac {\Delta{\bf R}^2}{4D_b(\nu)t}\,\right]\,\,\label{vgd}
\end{array}
\end{equation}
Hence, for right-hand sides of (\ref{inf}) and (\ref{inf0}) we obtain
estimates\,\, $\,\int_{n+1} V_{n+1}\,\sim\, \partial V_n/\partial \nu
\,\sim\,\nu^{-\,1}V_n\,$\, and
\begin{equation}
\begin{array}{l}
\int_1 V_1(t,{\bf R},{\bf r}_1,{\bf P},{\bf p}_1|{\bf
R}_0;\,\nu)\,\sim\, \frac {\partial V_0(t,\Delta{\bf R},{\bf
P}|\,\nu)}{\partial \nu}\,\sim\, \nu^{-\,1}\,\,V_0(t,\Delta{\bf
R},{\bf P}|\,\nu)\,\,\label{est}
\end{array}
\end{equation}
with coefficient $\,\nu^{-\,1}\,$ which is $\,(r_a^3\nu)^{-\,1}\gg
1\,$ times greater than in case of equilibrium correlations.

This observation prompts that characteristic volume, $\,\Omega_c\,$,
occupied by a historical correlation between BP and an atom by order
of magnitude is equal to $\,\nu^{-\,1}\,$,\, that is volume displayed
per one atom of fluid.

Physically, such statement follows also from almost trivial
syllogism:\,

(a) historical correlation between BP and atom springs from their
current (or soon forthcoming or just happening) collision;\,

(b) collision is (or will be or was) possible among those particles
only whose relative position vector $\,{\bf r}_1-{\bf R}\,$ belongs
to the ``collision cylinder''; the latter has radius $\,\approx
r_b\,$, is oriented along relative velocity of BP and atom, $\,{\bf
v}_1-{\bf V}_0={\bf p}_1/m-{\bf P}_0/M\,$, and its length in this
direction has an order of $\,\Lambda_b\,$ (assuming for simplicity
that $\,r_a\sim r_b\,$ and $\,\Lambda_a\sim\Lambda_b\,$) since
collision of stronger separated particles is prevented by other
particles;\,

(c) consequently, volume of the pair correlation $\,\Omega_c\sim \pi
r_b^2 \Lambda_b\sim\nu^{-\,1}\,$.

Formally, such statement, i.e.\, $\,\Omega_c\sim \nu^{-\,1}\,$,\,
follows from (\ref{est}) if one presumes that
\[
\begin{array}{l}
\max_{{\bf r}_1}\, |V_1(t,{\bf R},{\bf r}_1,{\bf P},{\bf p}_1|{\bf
R}_0;\,\nu)|\,\sim\, \left|\,\nu\,\frac {\partial V_0(t,\Delta{\bf
R},{\bf P}|\,\nu)}{\partial \nu}\right|G_m({\bf p}_1)\,\sim\,
\,V_0(t,\Delta{\bf
R},{\bf P}|\,\nu)\,\,G_m({\bf p}_1)\,\,\,,
\end{array}
\]
where maximum is achieved somewhere at\, $\,|{\bf r}_1-{\bf R}|\sim
r_b\,$.\, This presumption seems likely from the point of view of
equation (\ref{v1}) and well agrees with considerations of pair
correlations in conventional theory \cite{re,bal,dor}.

However, the resulting conclusion that \, $\,\Omega_c\sim
\nu^{-\,1}\,$\, contradicts conventional models of molecular
diffusion (``hybrid'' models, in terms of \cite{bal}, as they combine
dynamics and stochastics). In order to see this, let us consider
relation (\ref{cf1}) (in fact, definition of $\,V_1\,$) taking
$\,|{\bf r}_1-{\bf R}|>r_b\,$. Since left side of (\ref{cf1})
represents a probability density, it is certainly non-negative.
Therefore everywhere in the mentioned region, including most part of
the collision cylinder,
\begin{equation}
\begin{array}{l}
V_1(t,{\bf R},{\bf r}_1,{\bf P},{\bf p}_1|{\bf
R}_0;\,\nu)\,\,\geq\,\,-\,V_0(t,\Delta{\bf R},{\bf
P}|\,\nu)\,\,G_m({\bf p}_1)\,\,\label{in0}
\end{array}
\end{equation}
Hence, according to the estimate\, $\,\Omega_c\sim \nu^{-\,1}\,$,\,
\begin{equation}
\begin{array}{l}
\int_1 V_1(t,{\bf R},{\bf r}_1,{\bf P},{\bf p}_1|{\bf
R}_0;\,\nu)\,\,\gtrsim \,\, -\,\Omega_c\,V_0(t,\Delta{\bf R},{\bf
P}|\,\nu)\,\sim\, -\,\nu^{-\,1}\,\,V_0(t,\Delta{\bf R},{\bf
P}|\,\nu)\,\,\,,\label{in1}
\end{array}
\end{equation}
regardless of\, $\,t\,$ and $\,\Delta {\bf R}\,$. This inequality as
combined with (\ref{inf0}) or (\ref{est}) and confronted with
(\ref{vgd}) clearly shows that formula (\ref{vg}) can not represent a
true asymptotic of the BP's probability distribution!

A more rigorous substantiation of this striking statement was
suggested in \cite{jstat,last}. There (see also \cite{pro}) it was
shown that true asymptotic of $\,V_0(t,\Delta{\bf R},{\bf
P}|\,\nu)\,$ as a function of $\,\Delta{\bf R}\,$ possesses power-law
long tails cut off at $\,|\Delta{\bf R}|\sim v_b t\,$. This
qualitative prediction agrees with quantitative result of approximate
analysis of the BBGKY hierarchy undertaken in \cite{p1}\,:
\begin{equation}
V_0(t,\Delta {\bf R},{\bf P}|\,\nu)\,\rightarrow \,\,\frac {\Gamma
(7/2)}{[4\pi D_b(\nu )\, t\,]^{\,3/2}}\, \left[\,1+\frac {\Delta{\bf
R}^2}{4D_b(\nu )\,t}\,\right]^{-\,7/2}\Theta \left (\frac {|\Delta
{\bf R}|}{v_b t}\right )\,G_M({\bf P})\,\,\,,\label{as2}
\end{equation}
where function $\,\Theta(x)\approx 1\,$ at $\,x\ll 1\,$ and in a fast
way tends to zero at $\,x>1\,$. An origin of such statistics of
equilibrium molecular Brownian motion was discussed more than once as
long ago as in \cite{bk12,bk3} and later in \cite{i1,i2,p1,kmg} (see
also references therein). In short, such statistics says merely that
dynamics of a very many-particle system always is unique and can not
be imitated by primitive stochastic processes.

\section{Conclusion}

To resume, first, in this paper we rigorously formulated the problem
of random walk of small molecular-size Brownian particle (BP) in
classical thermodynamically equilibrium fluid. We considered
corresponding BBGKY hierarchy in terms of Bogolyubov type generating
functionals and derived a linear evolution equation for generating
functional of time-dependent ``historical  correlations'' which
together accumulate statistics of collisions of BP with fluid atoms
and eventually determine statistics of BP's path.

Second, we showed that both the evolution equation and its solution,
along with generating functional of static equilibrium correlations,
are invariant with respect to definite continuous group of
transformations of their independent variables including fluid
density. As the consequence, we found an infinite set of original
exact ``virial relations'' which connect the BP's path probability
distributions and various correlation functions at different values
of the density.

Notice that particular relation (\ref{vexp}) first was found in
\cite{pro} in the form integrated over BP's momentum and in
\cite{jstat,last} in the full form. There it was derived from most
general properties of the Liouville operator, as an example of the
``generalized fluctuation-dissipation relations'' \cite{j12,p}. Here,
formula (\ref{vexp}) appeared, in company with infinitely many new
relations (\ref{inf}) and (\ref{vexpn}), as consequence of hidden
symmetry of the BBGKY hierarchy. In principle, of course, these are
allied approaches. Nevertheless, they use very different formal
techniques, therefore confirmation of previous results certainly is
useful. The more so as it produces much wider results which give a
new sight of the BBGKY theory.

Third, we demonstrated that virial relations bring significant
information about solutions to the BBGKY hierarchy even without
literal solving it. In particular, they impose principal restrictions
on possible asymptotic profile of the BP's path probability
distribution, definitely forbidding the Gaussian profile implied by
invented stochastic or hybrid models of random walks
\cite{re,bal,dor}, in agreement with results previously obtained by
approximate methods in \cite{i1} (or see \cite{i2}) and in \cite{p1}.

Undoubtedly, this is not all of the benefits from the invariance
group of BBGKY hierarchy represented by virial relations. It can
extended, from one hand, to non-equilibrium random walks at presence
of an external force acting onto BP. From the other hand, to more
usual problems when probability distributions under interest describe
not information on position of some distinct particle but
hydrodynamic fields of initially non-equilibrium or externally
disturbed fluid. Curious feature of the virial relations
(\ref{vexpn})-(\ref{vexp}) is that they establish contacts between
states of a fluid under arbitrary different values of pressure, e.g.
between dense liquid and dilute gas. May be this will help in
constructing approximate solutions to the BBGKY equations. Anyway
there is a lot of interesting tasks for the future.

\end{document}